\documentclass[%
 reprint,
superscriptaddress,
 amsmath,amssymb,
 aps,
]{revtex4-2}

\usepackage{graphicx}% Include figure files
\usepackage{dcolumn}% Align table columns on decimal point
\usepackage{bm}% bold math
\usepackage{booktabs}%
\usepackage[english]{babel}
\usepackage{orcidlink} % Add ORCID links to authors
\usepackage{soul}

\newcommand\aap{A\&A}                % Astronomy and Astrophysics
\newcommand\apjl{ApJ}                % Astrophysical Journal, Letters
\newcommand\araa{ARA\&A}             % Annual Review of Astronomy and Astrophysics
\newcommand\jcap{J.~Cosmology Astropart. Phys.} % Journal of Cosmology and Astroparticle Physics
\newcommand\mnras{MNRAS}             % Monthly Notices of the Royal Astronomical Society
\newcommand\pasa{Publ. Astron. Soc. Australia}  % Publications of the Astronomical Society of Australia
\newcommand\ssr{Space Sci. Rev.}     % Space Science Reviews
\newcommand\arcmin{\hbox{$^\prime$}}
\newcommand\arcsec{\hbox{$^{\prime\prime}$}}

\newcommand{\IFAA}{Institute for Frontiers in Astronomy and Astrophysics, Beijing Normal University, Beijing 102206, China}
\newcommand{\SPA}{School of Physics and Astronomy, Beijing Normal University, Beijing 100875, China}
\newcommand{\SPLZL}{School of Physics and Laboratory of Zhongyuan Light, Zhengzhou University, Zhengzhou 450001, China}
\newcommand{\NAOC}{National Astronomical Observatories, Chinese Academy of Sciences, Beijing 100101, China}
\newcommand{\DAWU}{Department of Astronomy, Westlake University, Hangzhou 310030, China}
\newcommand{\IAAA}{Institute of Astronomy and Astrophysics, Anqing Normal University, Anqing 246133, China}

\begin{document}

% \maketitle

\title{Distinguishing the nature of dark matter by mapping cosmic filaments from Lyman-alpha emission}% Force line breaks with \\

\author{Yizhou Liu\orcidlink{0009-0005-8855-0748}}
% \email{liuyz.cosmo@gmail.com}
% \email{yzliu@bnu.edu.cn}
\affiliation{\IFAA}
\affiliation{\SPA}

\author{Liang Gao\orcidlink{0009-0006-3885-9728}}%
\email{lgao@bnu.edu.cn}
\affiliation{\IFAA}
\affiliation{\SPA}
\affiliation{\SPLZL}

\author{Shihong Liao\orcidlink{0000-0001-7075-6098}}%
\affiliation{\NAOC}

\author{Kai Zhu\orcidlink{0000-0002-2583-2669}}%
\affiliation{\DAWU}

\author{Yingjie Jing\orcidlink{0000-0003-3433-8416}}%
\affiliation{\NAOC}

\author{Huijie Hu\orcidlink{0000-0002-1908-0384}}%
\affiliation{\IAAA}
\affiliation{\NAOC}

\date{\today}% It is always \today, today,
             %  but any date may be explicitly specified

\begin{abstract}

The standard $\Lambda$CDM cosmological model predicts that cosmic filaments are highly clumpy, whereas warm dark matter -- invoked to address small-scale challenges in $\Lambda$CDM -- produces filaments that are noticeably smoother and less structured.  In this work, we investigate the potential of Lyman $\alpha$ (Ly$\alpha$) emission to trace cosmic filaments at redshifts $z=2.5$ and $z=4$, and assess their potential for constraining the nature of dark matter. Our analysis shows that Ly$\alpha$ filaments provide a promising observational probe of dark matter: at $z=4$, differences in filament smoothness and surface brightness serve as distinctive signatures between models. Looking ahead, the upcoming generation of 30-meter class telescopes will be critical for enabling these measurements, offering a compelling opportunity to distinguish the nature of dark matter by mapping the structure of cosmic filaments.

\end{abstract}

\maketitle

\section{\label{sec:level1}INTRODUCTION}

The standard cosmological model, $\Lambda$CDM, has been remarkably successful in explaining a diverse array of observational data, from the cosmic microwave background to the large-scale distribution of galaxies \cite{Springel2006}. In this paradigm, cosmic structures grow from primordial density fluctuations under the gravitational influence of dark matter -- a component whose fundamental nature remains unknown but which appears to interact solely through gravity \cite{Frenk2012}.  Dark matter is usually assumed to be ``cold", characterized by negligible thermal velocities, which allows the formation of dark matter halos with masses as small as that of Earth \cite{Hofmann2001, Green2004, Diemand2005, Wang2020, Liu2024, Zheng2024}.

Despite these successes, $\rm \Lambda$CDM faces persistent challenges on small scales, including the missing satellites, cusp-core, and too-big-to-fail problems \cite{Weinberg2015, Bullock2017, DelPopolo2017}. While baryonic processes such as cosmic reionization and stellar feedback can alleviate these discrepancies \cite{Governato2012, Pontzen2012, Zolotov2012, Brooks2014, Sales2022}, modifications to the dark matter paradigm offer an alternative explanation. Warm dark matter (WDM), such as sterile neutrino \cite{Dodelson1994, Dolgov2002},  with its non-negligible thermal velocities, suppresses small-scale structure formation and naturally mitigates some of these issues \cite{Colin2000, Avila-Reese2001, Bode2001, Schneider2012, Lovell2014, Ludlow2016, Bose2017, Lovell2017, Khimey2021, Paduroiu2022, Ussing2024}. However, given the degeneracy between baryonic effects and intrinsic dark matter properties, small-scale challenges alone are insufficient to conclusively validate or rule out the $\rm \Lambda$CDM framework.

The observational tools that are relatively independent of complicated baryonic uncertainties at galactic scales are therefore essential to break the degeneracy. The Ly$\alpha$ forest has been a powerful tool in this regard, constraining the one-dimensional power spectrum of the intergalactic medium \cite{Hernquist1996, Viel2004, Viel2005, Boyarsky2009, Viel2013, Baur2016, Irsic2017, Garzilli2019, Garzilli2021, Villasenor2023, Irsic2024}, which typically places lower limits at $3-5\,\mathrm{keV}$ \citep{Villasenor2023, Irsic2024}. However, the corresponding constraints remain subject to systematic uncertainties. In particular, degeneracies between the thermal history of the intergalactic medium and WDM properties can affect the inferred lower limits \citep{Garzilli2017, Garzilli2021}. In addition, differences in the treatment of low-density gas between hydrodynamical methods may affect the predicted Ly$\alpha$ flux power spectrum \citep{Chabanier2023}. These effects therefore indicate that the precise lower bound remains uncertain.

Gravitational lensing provides another complementary approach by directly probing substructure in dark matter halos \cite{Metcalf2001, Miranda2007, Zackrisson2010, Vegetti2012, Hezaveh2016_a, Hezaveh2016_b, Li2016, Minor2017, Gilman2020}. However, lensing constraints are also affected by uncertainties associated with halo modeling, line-of-sight structures, and observational systematics \citep{Despali2018, Vegetti2024}. Together, these methods have placed meaningful limits on dark matter models, but additional strategies are needed to further refine our understanding. 

An additional, independent, and intuitive probe is the distribution of neutral hydrogen (HI) within cosmic filaments. Prior work has shown that filamentary structures differ across dark matter models \cite{Knebe2003, Gao2007, Gao2015}. In particular, Gao et al. (2015) \cite{Gao2015} demonstrated that WDM yields smoother HI distributions in filaments than cold dark matter (CDM). If such differences can be detected via emission lines -- especially Ly$\alpha$ \cite{Elias2020, Witstok2021, Byrohl2023, Liu2025, Park2025} -- they would provide a direct and intuitive probe of dark matter physics. Recent reports of diffuse Ly$\alpha$ filaments \cite{Bacon2021, Bacon2023, Tornotti2025a, Tornotti2025b} further motivate a systematic investigation of how Ly$\alpha$ emission from filaments can be used to constrain the nature of dark matter.

In this work, we generate Ly$\alpha$ filament intensity maps at redshifts 2.5 and 4 using zoom-in hydrodynamical simulations of a Milky Way-like halo \cite{Gao2015}. The simulations are performed in two versions with the same random phases for the initial conditions: one assuming CDM and the other WDM. For the WDM scenario, we use a thermal relic mass of 1.5 keV, corresponding to a lower limit from satellite-count constraints \cite{Lovell2014}, to clearly highlight the differences between models, despite being disfavored by tighter Ly$\alpha$ forest constraints \cite{Villasenor2023, Irsic2024}. We then quantify the resulting differences in Ly$\alpha$ filament morphology and surface brightness between CDM and WDM, and discuss their implications for observationally probing the nature of dark matter.

\section{\label{sec:level2}METHODOLOGY}

\subsection{Simulation}
\label{sec:simulation} % used for referring to this section from elsewhere

We use two high-resolution zoom-in hydrodynamical simulations of a Milky Way-like galaxy \cite{Gao2015}, both run with {\small GADGET-3} \cite{Springel2005}. One assumes CDM and the other WDM. The simulations include cooling and photoheating of optically thin gas under a UV/X-ray background from Haardt et al. (2001) \cite{Haardt2001}. The adopted UV background model is not the most up-to-date and may introduce a bias in the Ly$\alpha$ predictions presented in later sections. However, we do not expect this uncertainty to qualitatively affect the relative comparison between the CDM and WDM simulations. As star formation efficiency is unclear in filaments, the simulations adopt a simple star formation criterion to convert a gas particle into a star when the particle's hydrogen density $n_{\rm H} > 0.1 \ \rm cm^{-3}$ and overdensity $\rho/\bar \rho > 2000$. A more physical treatment should consider the instability of the nearly one dimensional filament. However, the hydrodynamical simulations with radiative transfer show that the gas turns into mostly neutral above this threshold \citep{Altay2011, Rahmati2013}. 

In addition, hydrogen reionization is assumed to occur at $z=6$. The mass resolution is $5.16\times10^{4}\ h^{-1}M_{\odot}$ for gas and $2.35\times10^{5}\ h^{-1}M_{\odot}$ for dark matter. The simulations do not include feedback process. The re-simulated system is `halo A' from the Aquarius project \cite{Springel2008}, using the same cosmological parameters as previously stated. For the WDM run, the linear power spectrum is generated by truncating the CDM spectrum below the free-streaming scale, corresponding to a 1.5 keV thermal relic particle. Both simulations use identical random phases for the Gaussian initial conditions, ensuring a direct comparison between CDM and WDM. The adopted cosmological parameters are $\Omega_{\rm m}=0.25$, $\Omega_{\Lambda}=0.75$, $\sigma_{8}=0.9$, and $h=0.73$. While the adopted cosmological parameters deviate from the latest published values, the main conclusions of this work are qualitatively robust.

This work focuses on Ly$\alpha$ emission from gas in cosmic filaments. Since the simulations do not include self-shielding in dense filamentary gas, which strongly suppresses Ly$\alpha$ surface brightness, we correct for this by recomputing the neutral hydrogen ($n_{\rm HI}$), ionized hydrogen ($n_{\rm HII}$), and electron ($n_{\rm e}$) densities in each gas cell. Self-shielding is modeled following the suppression factor of Rahmati et al. (2013) \cite{Rahmati2013}. With this factor, we solve the ionization equilibrium equation to obtain corrected values of $n_{\rm HI}$, $n_{\rm HII}$, and $n_{\rm e}$.  

\begin{figure}[htbp]
\centering
% WDM 图片
\begin{minipage}[t]{0.5\textwidth}
    \centering
    \includegraphics[width=\linewidth]{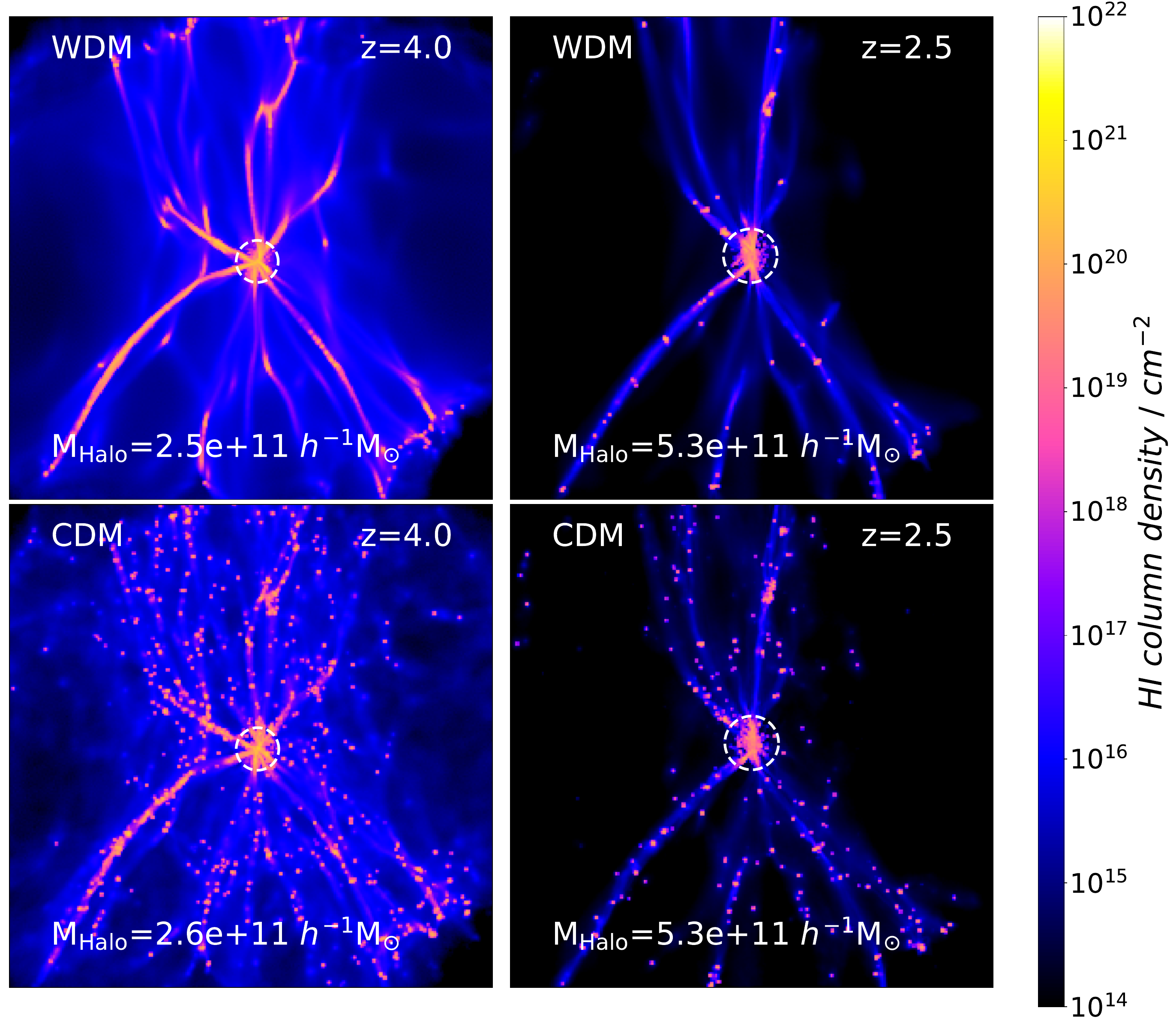}
    \caption{HI column density of warm and cold dark matter simulations at redshifts $z=2.5$ and $4$. The white circle marks the most massive halo at the center of the zoom-in region. Each panel spans 5.1 $\mathrm{cMpc}$.}
    \label{fig:temp and HI in wdm}
\end{minipage}
\hfill % 添加水平间距
% CDM 图片
\begin{minipage}[t]{0.5\textwidth}
    \centering
    \includegraphics[width=\linewidth]{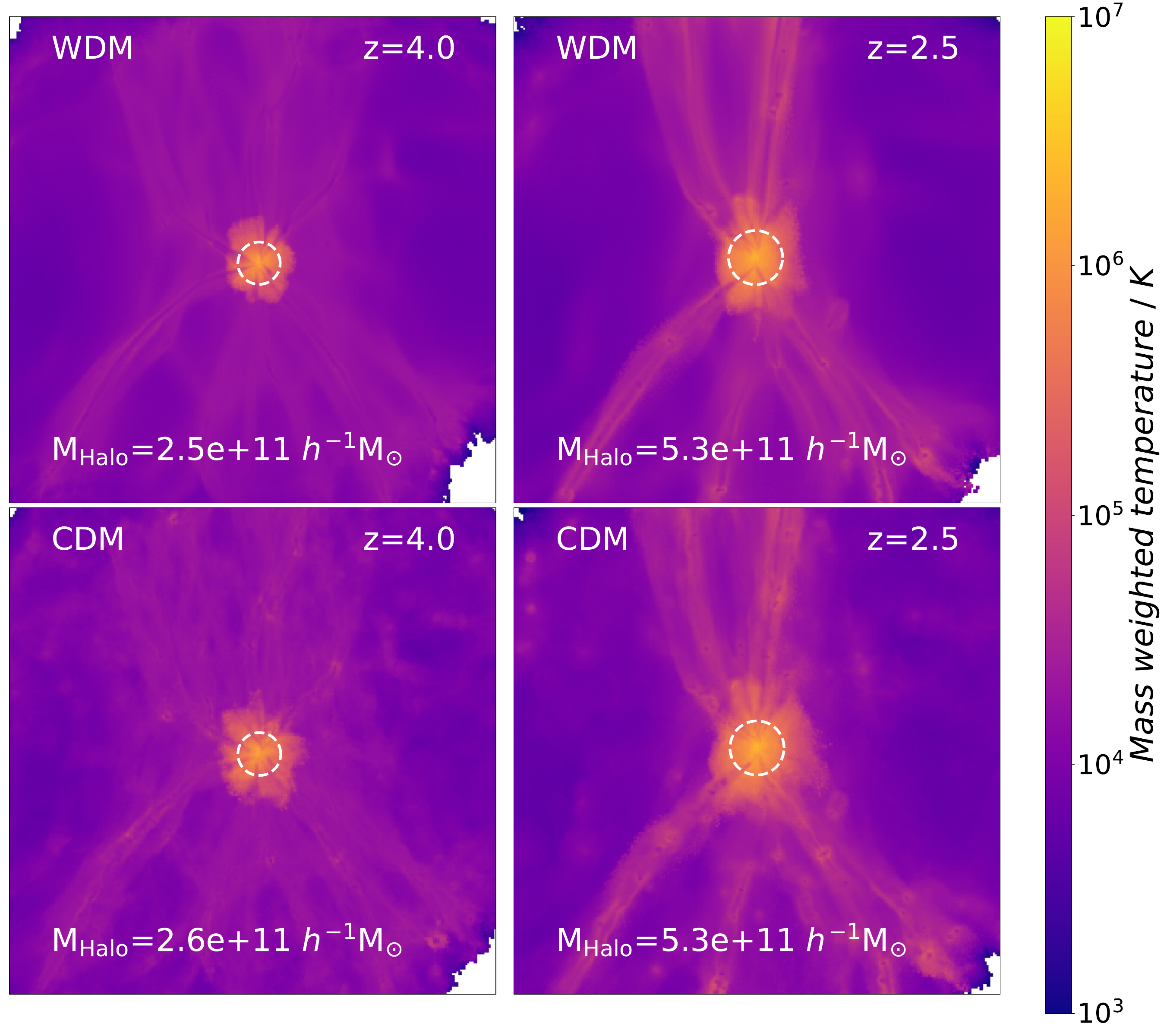}
    \caption{Same as Fig.~\ref{fig:temp and HI in wdm}, but for temperature maps.}
    \label{fig:temp and HI in cdm}
\end{minipage}
\end{figure}

\subsection{Ly$\alpha$ surface brightness map generation}

Ly$\alpha$ photons originate from the decay of hydrogen atoms from the 2p state to the ground state and are produced via two primary mechanisms. The first is recombination, given by:
\begin{equation}
    \epsilon_{\rm rec} = \alpha_{\rm B}(T)f_{\rm rec,B}(T)n_{\rm e}n_{\rm HII}E_{\rm Ly\alpha},
	\label{eq:recombination emission rate}
\end{equation}
where $\alpha_{\rm B}(T)$ and $f_{\rm rec,B}(T)$ are the temperature-dependence case B recombination coefficient and Ly$\alpha$ emission efficiency, respectively. $E_{\rm Ly\alpha}$ refers to the energy of a Ly$\alpha$ photon. The second mechanism is collisional excitation, expressed as:
\begin{equation}
    \epsilon_{\rm coll} = q_{\rm coll}(T)n_{\rm e}n_{\rm HI}E_{\rm Ly\alpha},
	\label{eq:collision emission rate}
\end{equation}
where $q_{\rm coll}(T)$ is the collisional excitation coefficient. 
The coefficients $\alpha_{\rm B}(T)$, $f_{\rm rec,B}(T)$, and $q_{\rm coll}(T)$ are adopted from \cite{Scholz1990, Scholz1991, Draine2011, Dijkstra2014, Silva2016}. 
The radiation transfer of Ly$\alpha$ photons is highly complex.
For further details and discussion of the modelling, we refer readers to Supplemental Material \S~\ref{sec: Emission Coefficients} and~\ref{sec: Simplified Radiative Transfer}.

\begin{table}[htbp]
 \caption{Mean transmission rates (Tr) at different redshifts.}
 \label{tab: the transmission rate at different redshift}
 \centering
 \begin{tabular}{p{1.5cm} | p{1.5cm} | p{1.5cm}}
  \toprule
  $z$ & 2.5 & 4\\
  \midrule
  Tr & 0.92 & 0.52\\
  \botrule
 \end{tabular}
\end{table}

To generate Ly$\alpha$ surface brightness maps, we first compute the total Ly$\alpha$ luminosity in each gas cell using its temperature, and number densities of neutral hydrogen, ionized hydrogen, and electrons. The luminosity is then projected onto a two-dimensional plane using the Cloud-In-Cell (CIC) method. A transmission factor is applied to account for line-of-sight attenuation, which is listed in Tab.~\ref{tab: the transmission rate at different redshift}. The value of transmission factor is recalculated according to the cosmological parameters adopted in our simulations, following Liu et al. (2025) \cite{Liu2025}. The calculation of this factor is described in the Supplemental Material \S~\ref{sec: Simplified Radiative Transfer}. Finally, the surface brightness map is obtained by converting the luminosity into a luminosity per unit area unit solid angle. 
 
\begin{figure}[htbp]
\centering
\includegraphics[width=0.5\textwidth]{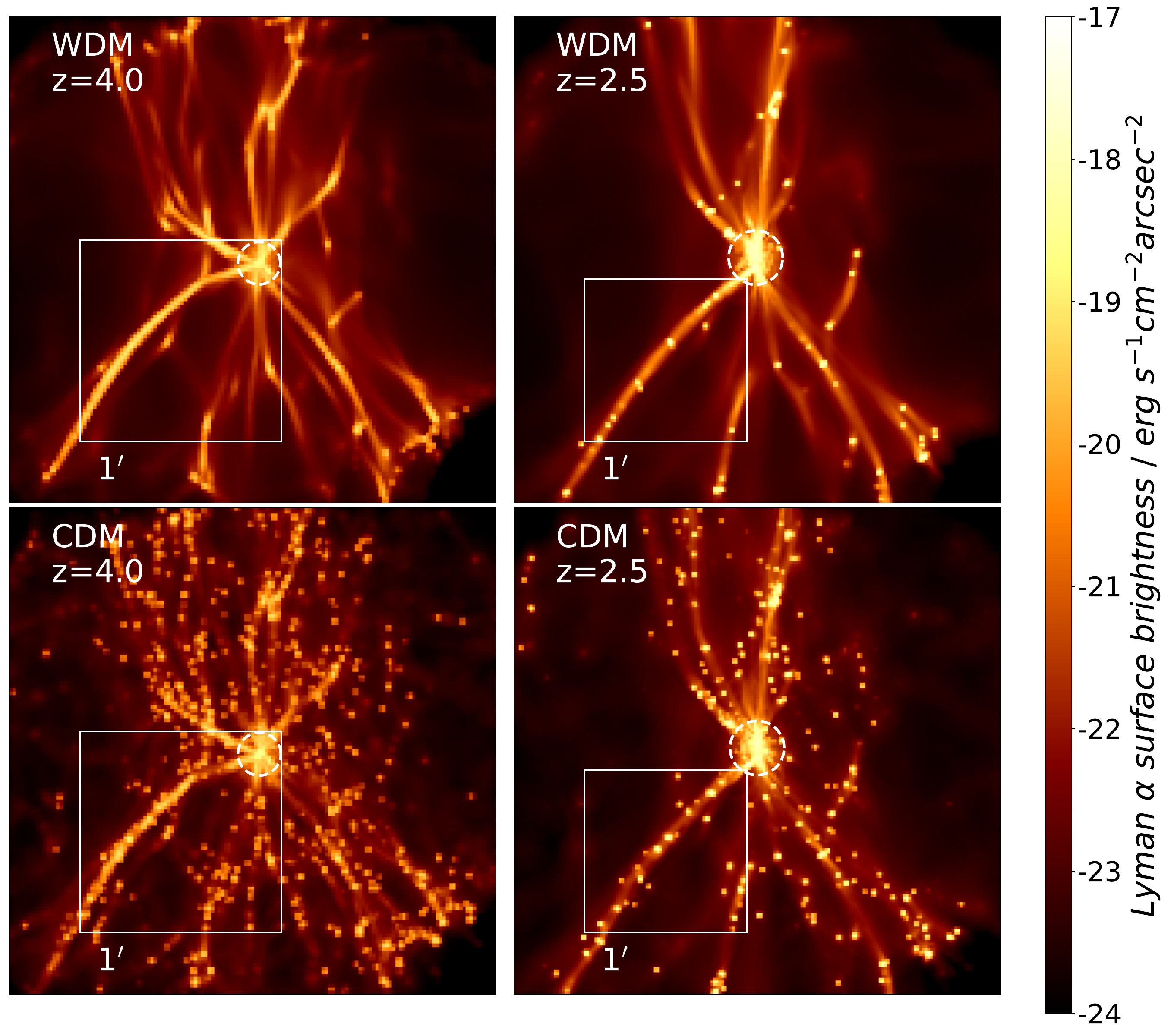}
\caption{The Ly$\alpha$ surface brightness of warm and cold dark matter simulations at redshifts $z=2.5$ and $4$. The white box represents the field-of-view of MUSE.}
\label{fig:lya of wdm and cdm}
\end{figure}

\begin{table*}[htbp]
\newcommand{\tabincell}[2]{\begin{tabular}{@{}#1@{}}#2\end{tabular}}
 \caption{Parameters of IFS.}
 \label{tab: the parameters of IFS}
 \centering
 \begin{tabular}{p{2.0cm} | p{2.0cm} | p{2.5cm} | p{2.5cm}}
  \toprule
  \tabincell{l}{IFS} & \tabincell{l}{FOV/\arcsec$^{2}$} & \tabincell{l}{Wavelength/nm} & \tabincell{l}{Redshift range} \\
  \midrule
  KCRM & $20\times33$ & 530-1050 & 3.36-7.64\\
  MUSE & $60\times60$ & 465-930 & 2.82-6.65\\
  MEGARA & $12.5\times11.3$ & 400-1000 & 2.29-7.23\\
  VIRUS & $50\times50$ & 350-550 & 1.88-3.52\\
  LRS2 & $12\times6$ & 370-1050 & 2.04-7.64\\
  \botrule
 \end{tabular}
\end{table*}

\section{\label{sec:level3}RESULTS}%

Fig.~\ref{fig:temp and HI in wdm} and~\ref{fig:temp and HI in cdm} show the HI column density and temperature maps of filaments in the warm and cold dark matter scenarios. The left and right columns correspond to redshifts $z=4$ and $2.5$, respectively, covering a co-moving region of about 5.1 cMpc. We highlight only the most massive halo at the center, whose mass ($\sim10^{11}\ h^{-1}\mathrm{M_{\odot}}$) is comparable to the Ly$\alpha$ emitter identified by Tornotti et al. (2025) \cite{Tornotti2025b} in the MUSE Ultra Deep Field (MUDF).

Cold streams flow along dark matter filaments and feed the central halo, fueling star formation in its galaxy. In our maps, five filaments connect to the central halo. Among them, the structure in the lower left corner is the longest ($\sim2$ cMpc) and shows the highest HI column density. The filaments are denser at higher redshift, consistent with our earlier findings \cite{Liu2025}. Notably, HI gas in the WDM case appears both denser and smoother than in CDM, especially at $z=4$. Cold gas follows a similar trend. These structural differences directly shape the Ly$\alpha$ appearance and surface brightness of the filaments.

Fig.~\ref{fig:lya of wdm and cdm} presents Ly$\alpha$ intensity maps for both models at $z=4$ and $2.5$, with a resolution of 1\arcsec. The narrowband width of slice is 6.25\,\AA\ at $z=4$ and 3.75\,\AA\ at $z=2.5$, respectively. The adopted narrowband widths are determined by the MUSE spectral resolution (1.25 \AA) and the thickness of the zoom-in regions along the line of sight. As expected, the contrast between CDM and WDM filaments is strongest at $z=4$ (left column), where the WDM filament is significantly smoother than its CDM counterpart. By $z=2.5$ (right column), small Ly$\alpha$ clumps begin to appear along the WDM filament, reducing the difference between models. This indicates that high-redshift Ly$\alpha$ filaments (e.g., at $z=4$) are more sensitive probes of dark matter properties, which implies a relatively narrow observational window for this probe. The cumulative distribution functions of the pixel surface-brightness values of Fig.~\ref{fig:lya of wdm and cdm} are presented in Supplemental Material \S~\ref{sec: Pixel Surface Brightness Statistics}. This statistic shows only a weak separation between the two models, except at the bright end, where the results are more sensitive to uncertain gas physics. This suggests that filament morphology may provide a more robust discriminator.

The integral field spectrographs (IFS) are considered to be well-suited instrument for detecting such faint and extended structures \cite{Witstok2021}. As shown in Liu et al. (2025) \cite{Liu2025}, Ly$\alpha$ filaments at $z \leq 4$ could be clearly detected with $\sim$30 m class telescopes equipped with a MUSE-like IFS. For these observations, the field of view (FOV) is more critical than angular resolution: the low surface brightness demands long exposure times, and a larger FOV increases the chance of capturing multiple and longer filaments within a reasonable integrating time. Tab.~\ref{tab: the parameters of IFS} lists instruments capable of detecting high-redshift filaments. Among them, MUSE offers the largest FOV ($1\arcmin\times1\arcmin$), shown by the white box in Fig.~\ref{fig:lya of wdm and cdm}, which is sufficient to capture filaments extending $1.5$-$2$ $h^{-1}$cMpc at $z=2.5$-$4$.

\begin{figure}[htbp]
\centering
\includegraphics[width=0.5\textwidth]{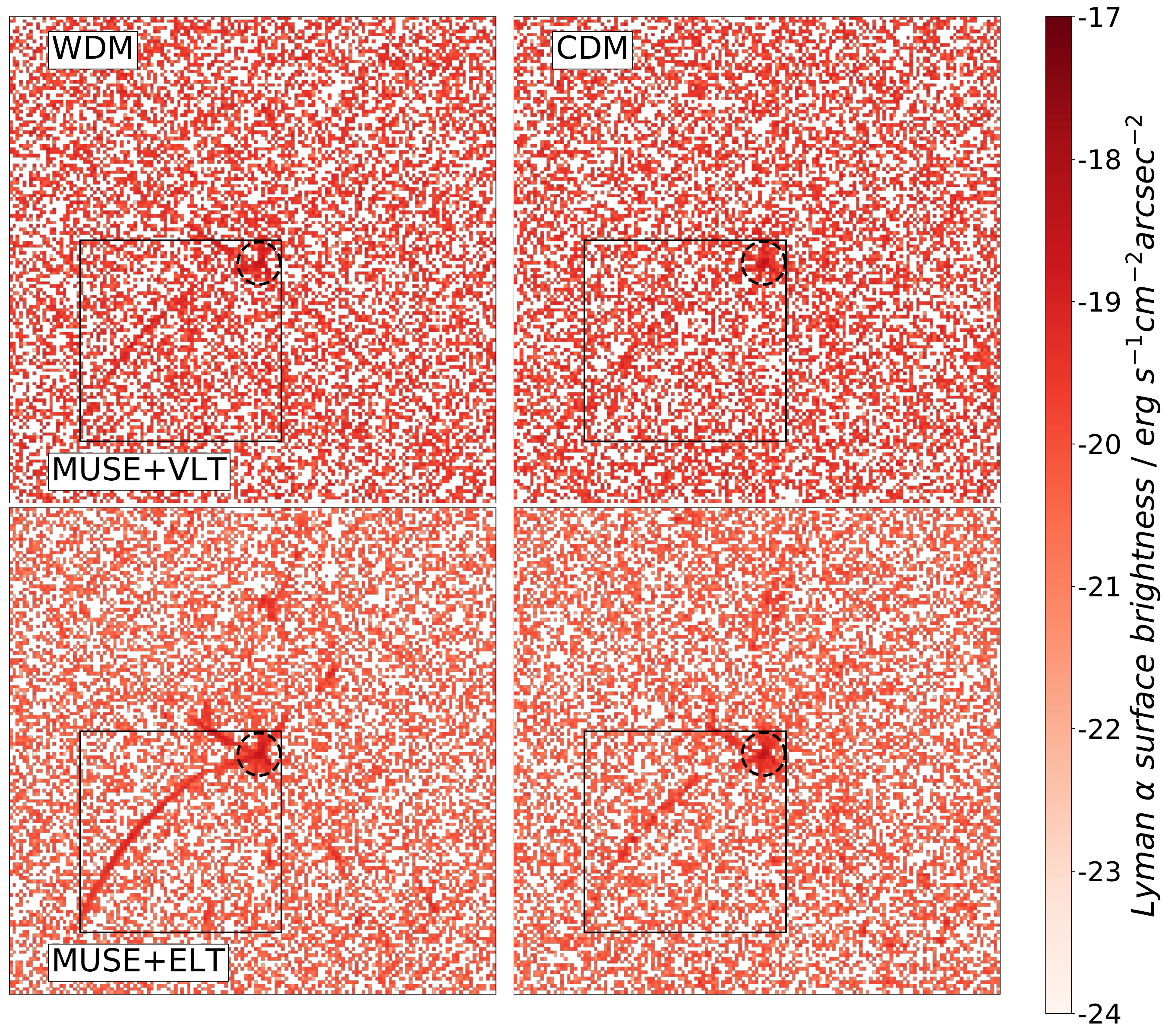}
\caption{Mock Ly$\alpha$ surface brightness maps at redshift $z=4$ for warm (left) and cold (right) dark matter. Rows correspond to the expected observational performance of different telescopes.}
\label{fig:lya of wdm and cdm with noise}
\end{figure}

For current $\sim$8m class telescopes, the main limitation is sensitivity. Liu et al. (2025) \cite{Liu2025} found that portions of filaments become detectable when the sensitivity reaches $\sim10^{-20}\ \rm erg\ s^{-1}cm^{-2}arcsec^{-2}$. To examine the thresholds required to distinguish dark matter models, we generated mock Ly$\alpha$ maps at $z=4$ for VLT and ELT (Fig.~\ref{fig:lya of wdm and cdm with noise}), adding Gaussian noise consistent with instrumental sensitivities (rescaled from Liu et al. 2025 \cite{Liu2025}, see Tab.~\ref{tab: the sensitivity of MUSE}). The rescaling method is described in the Supplemental Material \S~\ref{sec: Mock Observation Setup}. Each column shows a dark matter model, while rows correspond to telescopes.

Even with noise, WDM filaments remain smoother and brighter than CDM filaments. WDM filaments are marginally detectable with MUSE/VLT at noise levels of about $3.5\times10^{-20}$ $\rm erg\ s^{-1}cm^{-2}$ $\rm arcsec^{-2}$. In contrast, CDM filaments require higher sensitivity, achievable with MUSE-like instruments on the ELT, which can reach about $10^{-20}\ \rm erg\ s^{-1}cm^{-2}arcsec^{-2}$ -- again consistent with Liu et al. (2025) \cite{Liu2025}. However, reaching this sensitivity would require a substantial observational investment of approximately 150 hours on the ELT.

\begin{table}[htbp]
\newcommand{\tabincell}[2]{\begin{tabular}{@{}#1@{}}#2\end{tabular}}
 \caption{The sensitivity of MUSE installed on the VLT and ELT at $z=4$, for a bandwidth of 6.25 \AA\ and an exposure time of 150 hours. The unit of sensitivity is $10^{-20} \rm erg\ s^{-1}cm^{-2}arcsec^{-2}$.}
 \label{tab: the sensitivity of MUSE}
 \centering
 \begin{tabular}{p{2.5cm} | p{2.5cm}}
  \toprule
  Telescopes & Sensitivity \\
  \midrule
  VLT & 3.5 \\
  \midrule
  ELT & 0.8 \\
  \botrule
 \end{tabular}
\end{table}

These findings suggest that both the smoothness and surface brightness of Ly$\alpha$ filaments can serve as diagnostic features for distinguishing between dark matter models. 
Among them, filament morphology may provide the more robust diagnostic.
Current surveys may already be able to look for smooth, extended filaments indicative of WDM, such as MUDF \cite{Tornotti2025b} and MXDF \cite{Bacon2021, Bacon2023}. However, the rarity of such detections limits strong conclusions. 

Note that the above results are based on a high-resolution simulation of a single halo. Halo-to-halo variance and viewing geometry may affect the detailed appearance of the filaments. A statistical assessment using a larger halo sample will therefore be important in future work. Additionally, quantitative measures beyond visual comparison, such as the power spectrum, clumping factor, and structure function, should also be explored.

Finally, we also note that our analysis considers only a WDM  model with a particle mass of $1.5\,\mathrm{keV}$, a lower bound from satellite-count constraints \cite{Lovell2014}. Larger WDM particle mass would reduce the suppression of small-scale structure and thus diminish the differences between WDM and CDM filaments. Identifying the mass scale at which the two models become indistinguishable in Ly$\alpha$ emission requires a dedicated suite of simulations with varying WDM particle masses and is left to future work. 

\section{\label{sec:level4}DISCUSSION and CONCLUSION}%

This work presents a proof-of-concept demonstration of Ly$\alpha$ filaments as an independent probe of dark matter, however, there are some limitations in our modelling. To place these results in context, we now discuss the key modelling assumptions underlying our analysis and assess how they may influence the robustness of our conclusions. As illustrated in the Methods section, the predictions presented in this work rely on a few modelling assumptions that are common to current studies of Ly$\alpha$ emission from large-scale structure. Here we briefly summarise the most relevant assumptions and assess their potential impact on our conclusions.

First, the Ly$\alpha$ emission from cosmic filaments is modelled using a simplified but physically motivated prescription. As described in detail in our previous work \cite{Liu2025} and Supplemental Material, this model is constructed to provide a conservative estimate of the intrinsic Ly$\alpha$ surface brightness, intentionally yielding a lower bound rather than an optimistic prediction. As a result, the filament detectability inferred in this study is unlikely to be overestimated due to uncertainties in the emission modelling.  
However, although the full Ly$\alpha$ radiative-transfer effect is relatively weak in the dense regions that dominate filament emission and mainly affects the diffuse outskirts \cite{Byrohl2023, Liu2025}, it still could reduce the power of this probe slightly.

A second important source of uncertainty arises from baryonic feedback processes in warm dark matter simulations. As noted in the Methods, the simulations analysed here do not include explicit feedback. Feedback could either reduce or enhance the contrast between CDM and WDM. It may reduce the contrast by dispersing dense clumps and smoothing the gas distribution in CDM. Conversely, it may enhance the contrast, since star formation is expected to be more clustered in CDM owing to its larger abundance of low-mass halos. This could lead to stronger gas heating and redistribution in CDM filaments, making them more inhomogeneous. In WDM, by contrast, the smaller number of stars forming along filamentary structures may drive feedback preferentially perpendicular to the filament axis, potentially yielding a more regular gas distribution. The net effect remains uncertain and should be assessed with future simulations including more complete baryonic physics.

In the near future, the increased sensitivity and larger collecting areas of $\sim$30m class telescopes will allow more robust measurements of high-redshift Ly$\alpha$ filaments, potentially providing a promising and independent avenue for constraining dark matter models. Although the above results are derived from WDM simulations, they should also be applicable to the fuzzy dark matter (FDM) model, which exhibits similar filamentary structures due to a comparable suppression of small-scale density fluctuations \cite{Mocz2019, May2021, May2023, Nori2023, Zimmermann2025}. However, FDM filaments are additionally shaped by wave effects and interference patterns that are absent in thermal WDM. Ly$\alpha$ filaments may therefore offer a potential probe of FDM-specific signatures, although dedicated simulations are required to distinguish them from WDM and CDM predictions.

\begin{acknowledgments}

We are grateful to the anonymous referees for their constructive and insightful reports, which have significantly improved this work. We would like to thank Davide Tornotti, Joris Witstok, Ewald Puchwein, and Chris Byrohl for their help and discussions. We are also grateful to Simon D. M. White and Zheng Zheng for their helpful discussions and comments. We acknowledge the supports from the National Natural Science Foundation of China (Grant No. 12588202) and the National Key Research and Development Program of China (Grant No. 2023YFB3002500). H.H. is supported by the NSFC Grant Nos. 12503012.

\end{acknowledgments}

% The \nocite command causes all entries in a bibliography to be printed out
% whether or not they are actually referenced in the text. This is appropriate
% for the sample file to show the different styles of references, but authors
% most likely will not want to use it.
\nocite{*}

\bibliography{apssamp}% Produces the bibliography via BibTeX.

%%%%%%%%%%%%%%%%%%%%%%%%%%%%%%
%%% Supplementary material %%%
%%%%%%%%%%%%%%%%%%%%%%%%%%%%%%
% \clearpage
\onecolumngrid
\newpage
\begin{center}
    \textbf{\large{\MakeUppercase{Supplemental Material}}} \\
\end{center}
\twocolumngrid

% 取消后续章节编号
\setcounter{section}{0}
\setcounter{equation}{0}
\setcounter{figure}{0}
\setcounter{table}{0}
\setcounter{page}{1}

\section{Ly$\alpha$ Surface Brightness Modelling}

We summarize here the key emission coefficients, simplified radiative transfer, and mock observation setup underlying the Ly$\alpha$ surface brightness maps, focusing on aspects not explicitly described in the main text.

\subsection{Emission Coefficients} \label{sec: Emission Coefficients}

While the main text presents the recombination and collisional excitation rates, the specific formulae for the coefficients are provided here for reproducibility:

For coefficients of recombination:
\begin{equation}
\begin{split}
\alpha_{\rm B}(T) = a_{1}
\left(\frac{T}{10^{4}\,{\rm K}}\right)^{-a_{2}-a_{3}\log_{10}\left(\frac{T}{10^{4}\,{\rm K}}\right)}
{\rm cm^{3}\,s^{-1}},
\end{split}
\end{equation}
where $a_{1} = 2.54\times10^{-13}$, $a_{2} = 0.8163$, $a_{3} = 0.0208$ \cite{Draine2011}.
\begin{equation}
\begin{split}
f_{\rm rec,B}(T)=b_{1}-b_{2}\log_{10}\left(\frac{T}{10^{4}\,{\rm K}}\right)-b_{3}\left(\frac{T}{10^{4}\,{\rm K}}\right)^{-b_{4}},
\end{split}
\end{equation}
where $b_{1} = 0.686$, $b_{2} = 0.106$, $b_{3} = 0.009$, $b_{4}=0.44$ \cite{Dijkstra2014}.

For coefficients of collisional excitation:
\begin{equation}
q_{\rm coll}(T)=\gamma(T)\exp\left(-\frac{E_{\rm Ly\alpha}}{k_{\rm B}T}\right),
\end{equation}
where $k_{\rm B}$ is the Boltzmann constant and
\begin{equation}
\gamma(T)=\exp\left[\sum_{i=0}^{5}c_i\left(\ln T\right)^i\right].
\end{equation}
The parameters $c_{i}$ are listed in Tab.~\ref{tab: collisional_coefficients} \cite{Scholz1990, Scholz1991}.

\begin{table*}[htbp]
\newcommand{\tabincell}[2]{\begin{tabular}{@{}#1@{}}#2\end{tabular}}
 \caption{Polynomial coefficients used in the collisional excitation coefficient
$q_{\rm coll}(T)$.}
 \label{tab: collisional_coefficients}
 \centering
 \begin{tabular}{p{3.5cm} | p{2.0cm} | p{2.0cm} | p{2.0cm} |  p{2.0cm} |  p{2.5cm} |  p{2.5cm}}
  \toprule
  \tabincell{l}{Temperature range (K)} & \tabincell{l}{$c_0$} & \tabincell{l}{$c_1$} & \tabincell{l}{$c_2$} & \tabincell{l}{$c_3$} & \tabincell{l}{$c_4$} & \tabincell{l}{$c_5$} \\
  \midrule
  $2\times10^{3}\le T<6\times10^{4}$ & $-163.0155$ & $87.95711$ & $-20.57117$ & $2.359573$ & $-0.1339059$ & $3.021507\times10^{-3}$ \\
    
  $6\times10^{4}\le T<6\times10^{6}$ & $527.9996$ & $-193.9399$ & $27.18982$ & $-1.883399$ & $6.462462\times10^{-2}$ & $-8.811076\times10^{-4}$   \\

  $6\times10^{6}\le T<10^{8}$ & $-2813.3632$ & $815.09685$ & $-94.418414$ & $5.4280565$ & $-0.1546712$ & $1.7439112\times10^{-3}$\\
  \botrule
 \end{tabular}
\end{table*}

These coefficients, together with the gas temperature and densities, determine the intrinsic Ly$\alpha$ emissivity in each simulation cell.

\subsection{Simplified Radiative Transfer} \label{sec: Simplified Radiative Transfer}

The radiative transfer of Ly$\alpha$ photons is intrinsically complex, involving resonant scattering, frequency diffusion, and absorption by dust and molecular hydrogen \cite{Dijkstra2014}. These processes can significantly modify the spatial distribution and observed surface brightness of Ly$\alpha$ emission. A full treatment therefore requires dedicated Ly$\alpha$ radiative-transfer calculations, which are beyond the scope of the present work.

In this study, we focus on the intrinsic Ly$\alpha$ emission produced by diffuse filament gas. Ly$\alpha$ photons originating from star-forming gas are not included, as gas above the adopted star-formation threshold is converted into star particles in the simulations. We therefore neglect resonant scattering and dust absorption, and only account for the attenuation of Ly$\alpha$ photons along the line of sight through the intergalactic medium (IGM).

To model this effect, we apply a mean transmission factor, $\rm Tr$, to the intrinsic Ly$\alpha$ luminosity. We first employ Monte Carlo code to determine the emergent spectra of Ly$\alpha$ photons escaping from dense filament gas. The propagation of these photons through the IGM is then modeled assuming a spatially averaged HI distribution in ionization equilibrium with the adopted UV background and cosmological parameters. Under this assumption, the neutral hydrogen fraction and corresponding optical depth along the line of sight can be estimated, yielding an effective transmission factor $\rm Tr=e^{-\tau}$. The resulting values at $z=2.5$ and $z=4$ are listed in Tab.~\ref{tab: the transmission rate at different redshift} of the main text.

It should be noted that both Ly$\alpha$ photons escaping from galaxies and resonant scattering generally act to increase the spatial extent and total observable emission associated with cosmic filaments. By neglecting these contributions, our calculation is expected to provide a conservative lower estimate of the observable Ly$\alpha$ surface brightness. 

\subsection{Mock Observation Setup} \label{sec: Mock Observation Setup}

The procedure used to generate Ly$\alpha$ surface-brightness maps is described in the main text. Here we summarize the observational assumptions adopted in the mock observations. The thickness of each projected slice is chosen to match the narrowband width of a realistic integral-field spectroscopic observation. The observed narrowband width is taken as an integer multiple of the MUSE spectral resolution element (1.25\,\AA). The number of wavelength slices is selected such that the corresponding comoving depth remains comparable to the thickness of the zoom-in region along the line of sight. At $z=4$, we adopt 5 spectral slices, corresponding to a narrowband width of 6.25\,\AA, while it is selected as 3 spectral slices at $z=2.5$, corresponding to a narrowband width of 3.75\,\AA. The fiducial observational parameters are based on the MXDF survey \citep{Bacon2021}, including the angular resolution and sensitivity. The projected pixel size is determined from the adopted angular resolution 1\arcsec.

For the mock observations shown in Fig.~\ref{fig:lya of wdm and cdm with noise} of main text, Gaussian noise is added to each pixel according to the expected instrumental sensitivity. The reference noise level of MUSE/VLT is taken from the MXDF observations \cite{Bacon2021}. The corresponding values, together with the rescaled MUSE/ELT values, are listed in Table 3 of \citet{Liu2025}. For the different numbers of spectral slices adopted in this work, the noise level is rescaled as
\begin{equation}
\sigma_{\rm SB} = \sigma_{\rm ref}\left(\frac{N_{\rm SB}}{N_{\rm ref}}\right)^{1/2},
\end{equation}
where $\sigma_{\rm ref}$ is the reference noise level, and $N$ denotes the number of spectral slices.

\section{Pixel Surface Brightness Statistics}
\label{sec: Pixel Surface Brightness Statistics}

\begin{figure}[htbp]
\centering
\includegraphics[width=0.5\textwidth]{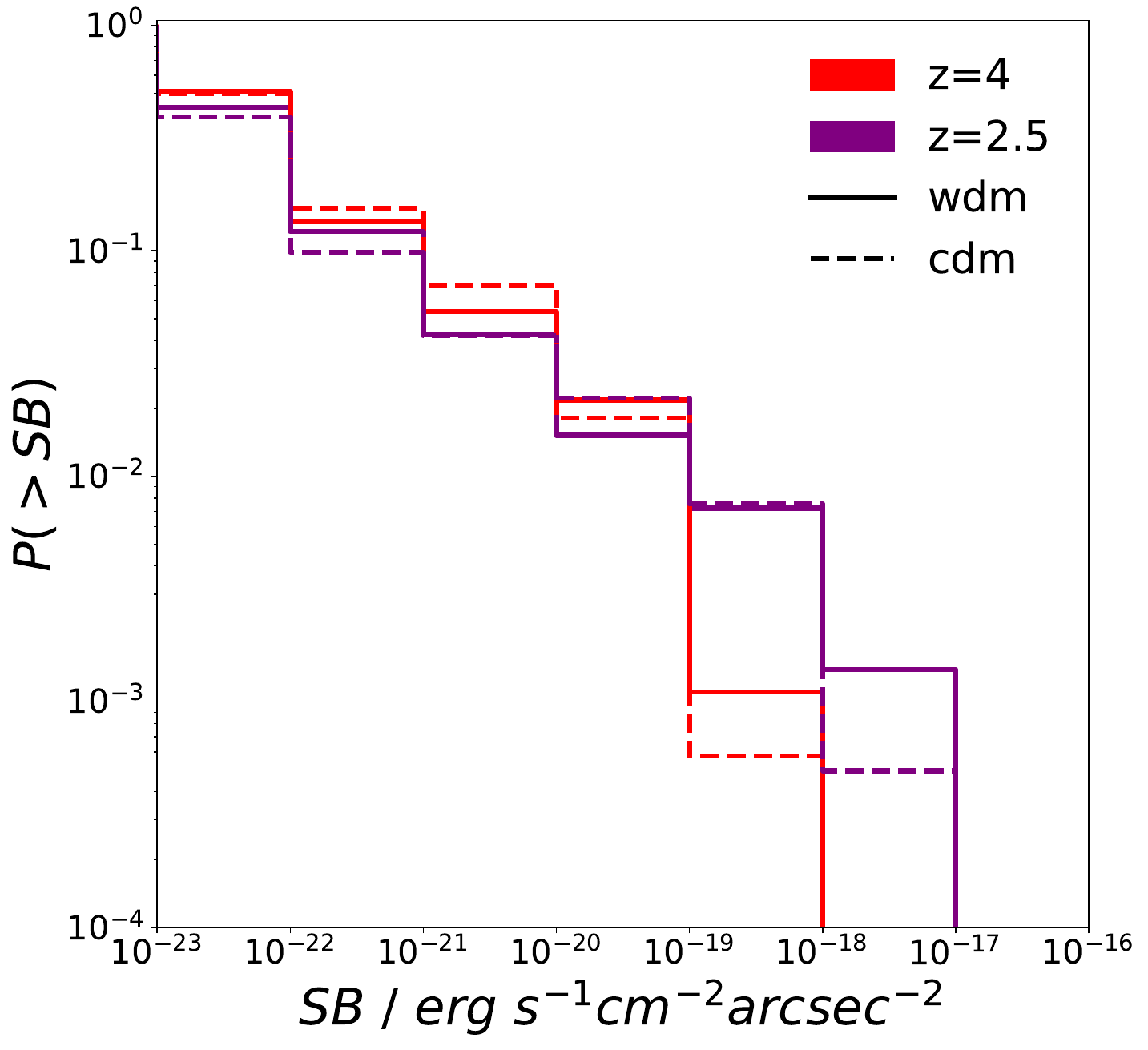}
\caption{The cumulative distribution functions of pixel surface-brightness values for the Ly$\alpha$ maps shown in Fig.~\ref{fig:lya of wdm and cdm} of main text. Red and purple lines indicate $z=4$ and $z=2.5$, respectively, while dashed and solid lines represent the CDM and WDM models.}
\label{fig:lya stats}
\end{figure}

In Fig.~\ref{fig:lya stats}, we show the cumulative distribution functions (CDFs) of the pixel surface-brightness values derived from the Ly$\alpha$ maps in Fig.~\ref{fig:lya of wdm and cdm} of the main text. Different colors correspond to different redshifts, while different line styles denote the CDM and WDM models. This figure indicates that the two models yield largely consistent results, except at the bright end, where gas-physics uncertainties become more important. We therefore consider the morphological smoothness and regularity of filaments to be a more robust discriminator, while this statistic provides a useful quantitative complement to the visual comparison.

\clearpage

\end{document}